\documentclass[12pt]{article}
\usepackage{psfrag}
\usepackage{graphicx}
\usepackage{graphics}
\newcommand{\As}{A\!\!\!/}
\newcommand{\ks}{k\!\!\!/}

\date{}
\begin{document}
\baselineskip=18.6pt plus 0.2pt minus 0.1pt \makeatletter


\title{\vspace{-3cm}
\hfill\parbox{4cm}{\normalsize \emph{LPHEA 04-09}}\\
 \vspace{1cm}
{A note on the polarization of the laser field in Mott
scattering}}
 \vspace{2cm}

\author{Y.  Attaourti$^{\dag}$\thanks{attaourti@ucam.ac.ma}, B. Manaut$^\ddag$ and S. Taj$^\ddag$\\
 {\it {\small $^\dag$Laboratoire de Physique des Hautes
Energies et d'Astrophysique, Facult\'e }}\\ {\it {\small des
Sciences Semlalia, Universit\'e Cadi Ayyad Marrakech, BP : 2390,
Maroc.}}
\\
{\it {\small $^\ddag$UFR de Physique Atomique Mol\'eculaire et
Optique Appliqu\'ee,
 Facult\'e des Sciences,}}\\
{\it {\small Universit\'e Moulay Isma\"{\i}l
   BP : 4010, Beni M'hamed, Mekn\`es, Maroc.}}\\
   {\it {\small }}}\maketitle \setcounter{page}{1}
\begin{abstract}
In the first Born approximation and using an elliptically
polarized laser field, the Mott scattering of an electron by a
Coulomb potential is investigated using the Dirac-Volkov states to
describe the incident and scattered electrons. The results
obtained are compared with the results of S.M. Li \textit{et al}
\cite{1} for the case of a linearly polarized laser field and with
the results of Y. Attaourti \textit{et al} \cite{2} for the case
of a circular polarization.
\\ \vspace{.04cm}\\
 PACS number(s): 34.80.Qb, 12.20.Ds
\end{abstract}

\maketitle
\section{Introduction}

In this note, we give definite analytical results concerning the
process of Mott scattering in the presence of a strong laser field
and compare our results to previous theoretical works, namely the
work of S.M. Li \textit{et al} \cite{1} and the work of Y.
Attaourti \textit{et al} \cite{2}. We hope this contribution will
be useful to all researchers working in this field and will bring
an end to the controversy raised by the expression found by C.
Szymanowski \textit{et al} \cite{3} for the case of a circular
polarization of the laser field. An analytical expression for the
spin-unpolarized differential cross section is derived using trace
calculations. The electric field strength as well as the frequency
of the laser field and the kinetic energy of the incoming electron
being key parameters, the study of the process of Mott scattering
in the presence of an elliptically polarized laser field
introduces a new key parameter, namely the degree of ellipticity
$\eta $. The cross section dependency on this new key parameter is
reported. The general features of the Mott scattering process are
qualitatively modified when a laser field is present and this is
particularly true when one study the spin-dependent relativistic
Mott scattering. Not only it is important to take care of the fact
that the electron is a fermion but also to describe this particle
by the appropriate wave function in a non-perturbative way. This
is done by using the Dirac-Volkov wave functions \cite{4} which
contain the interaction of the electron with the laser field to
all orders. The organization of this paper is as follows. In
section II, we present the theory in the first Born approximation.
In section III, we discuss the analytical results for the
spin-unpolarized differential cross section modified by the laser
field and analyze their dependencies on the new relevant
parameter, that is the degree of ellipticity $\eta $. We end by a
brief summary and conclusion in section IV. Throughout this work,
we use atomic units $\hbar =m=e=1$ and work with the metric tensor
$g^{\mu \nu }=diag(1,-1,-1,-1)$.

\section{Theory}

We treat the laser field classically since we are considering
intensities that do not allow pair creation \cite{3}. The four
potential corresponding to the laser field satisfies the Lorentz
condition $\partial ^{\mu }A_{\mu }=0$ and is given by

\begin{equation}
A=a_{1}\cos (\phi )+a_{2}\sin (\phi )\tan (\eta /2) , \label{1}
\end{equation}
with $\phi =k.x=k^{\mu }x_{\mu }=wt-\mathbf{k.x}$ and where $\eta
$ is the degree of ellipticity of the laser field. The four
vectors $a_{1}$ and $a_{2}$ satisfy the following relations
$a_{1}^{2}$ $=a_{2}^{2}=a^{2}$ and the Lorentz condition $k.A$
implies $a_{1}.k=a_{2}.k=0$. The linear polarization
is obtained for $\eta =0$ and the circular polarization is obtained for $%
\eta =\pi /2$. The electric field associated with the potential of
the laser field is

\begin{equation}
\mathbf{E}=-\frac{1}{c}\frac{\partial }{\partial t}\mathbf{A}
.\label{2}
\end{equation}

\subsection{Differential cross section}

The interaction potential is the Coulomb potential of a target
nucleus of charge $Z$

\begin{equation}
A_{Coul}^{\mu }=(-\frac{Z}{\mid \mathbf{x}\mid },0,0,0) ,
\label{3}
\end{equation}
and in the first Born approximation, the transition matrix element
for the transition ($i\rightarrow f$) is

\begin{equation}
S_{fi}=\frac{iZ}{c}\int d^{4}x\overline{\psi }_{q_{f}}(x)\frac{\gamma ^{0}}{%
\mid \mathbf{x}\mid }\psi _{q_{i}}(x)  .\label{4}
\end{equation}
The Dirac-Volkov wave functions $\psi _{q_{i}}(x)$ and $\psi
_{q_{f}}(x)$ describe the incident and scattered electron
respectively. Such wave functions normalized to the volume $V$ are
\cite{4}
\begin{equation}
\psi _{q}(x)=R(q)\frac{u(p,s)}{\sqrt{2QV}}e^{iS(q,x)},  \label{5}
\end{equation}
where $u(p,s)$ represents a Dirac bispinor normalized as $\overline{u}%
(p,s)u(p,s)=2c^{2}$ and $q^{\mu }=(Q/c,\mathbf{q})$ is the
quasi-impulsion acquired by the electron in the presence of the
laser field
\begin{equation}
q^{\mu }=p^{\mu }-\frac{1}{2c^{2}(k.q)}\overline{A^{2}}k^{\mu }
.\label{6}
\end{equation}
The quantity $R(q)$ is defined by $R(q) =1+\ks\As/(2c(k.q)) $, and
$S(q,x)=-(q.x)-((a_1.q)\sin(\phi)-(a_2.q)\cos(\phi)\tan(\eta/2))/(c(k.q))$,
where the Feynmann slash notation is used \cite{5} : for a given
four vector $A$, we have $\As=\gamma ^{\mu }A_{\mu }$. Finally the
averaged squared potential $\overline{A^{2}}$ is given by

\begin{equation}
\overline{A^{2}}=a^{2}(1+\tan^2(\eta /2))/2 , \label{7}
\end{equation}
from which one deduces $\overline{A^{2}}=a^{2}$ for the case of a
circular polarization of the laser field and
$\overline{A^{2}}=a^{2}/2$ for the case of a linear polarization
of the laser field. The argument $z$ of the ordinary Bessel
functions is

\begin{equation}
z=\sqrt{\alpha _{1}^{2}+\alpha _{2}^{2}}  ,\label{8}
\end{equation}
with

\begin{equation}
\alpha
_{1}=\frac{(a_{1}.p_{i})}{c(k.p_{i})}-\frac{(a_{1}.p_{f})}{c(k.p_{f})}
,\label{9}
\end{equation}
and
\begin{equation}
\alpha _{2}=[\frac{(a_{2}.p_{i})}{c(k.p_{i})}-\frac{(a_{2}.p_{f})}{c(k.p_{f})%
}]\tan (\eta /2) . \label{10}
\end{equation}
A useful parameter that intervenes in the expression of the DCS is
$\phi_0=\arctan(\alpha_2/\alpha_1)$.
 Using the standard procedures of QED \cite{5}, we obtain for the
spin-unpolarized differential cross section evaluated for
$Q_{f}=Q_{i}+sw$

\begin{equation}
\frac{d\overline{\sigma }}{d\Omega _{f}}=\sum_{s=-\infty }^{s=\infty }\frac{d%
\overline{\sigma }^{(s)}}{d\Omega _{f}},  \label{11}
\end{equation}
with
\begin{equation}
\frac{d\overline{\sigma }^{(s)}}{d\Omega _{f}}=\frac{Z^2
\mid \mathbf{q}_{f}\mid }{c^{4}\mid \mathbf{q}_{i}\mid }\frac{1}{|\mathbf{q}_f-\mathbf{q}_i-s\mathbf{k}|^4}\Big(\frac{1}{2}%
\sum_{s_{i},s_{f}}\mid M_{fi}^{(s)}\mid ^{2}\Big) . \label{12}
\end{equation}
Using REDUCE for the trace calculations \cite{6}, we obtain
\begin{eqnarray}
&&\frac{1}{2}\sum_{s_{i},s_{f}} \mid M_{fi}^{(s)}\mid
^{2}=2\Big\{J_{s}^{2}(z)A+(J_{s+1}^{2}(z)+J_{s-1}^{2}(z))B
\nonumber
\\
&&+J_{s+1}(z)J_{s-1}(z)C+J_{s}(z)(J_{s+1}(z)+J_{s-1}(z))D\nonumber
\\
&&+J_{s}(z)(J_{s+2}(z)+J_{s-2}(z))E+(J_{s+2}^{2}(z)+J_{s-2}^{2}(z))F\nonumber
\\ &&+(J_{s-1}(z)J_{s+2}(z)+J_{s+1}(z)J_{s-2}(z))G \nonumber \\
&&+(J_{s+1}(z)J_{s+2}(z)+J_{s-1}(z)J_{s-2}(z))H\Big\}.  \label{13}
\end{eqnarray}
The eight coefficients $A$, $B$, $C$, $D$, $E$, $F$, $G$ and $H$
are given respectively by
\begin{eqnarray}
A&=&c^{4}-(q_{i}.q_{f})c^{2}+2Q_{i}Q_{f}-\frac{a^{2}}{2}\Big(\frac{(k.q_{f})}{
(k.q_{i})}+\frac{(k.q_{i})}{(k.q_{f})}\Big)  \nonumber \\
&+&\frac{a^{2}w^{2}}{c^{2}(k.q_{i})(k.q_{f})}((q_{i}.q_{f})-c^{2})(1+\tan^2
(\eta /2))/2  \nonumber \\
&+&\frac{(a^{2})^{2}w^{2}}{c^{4}(k.q_{i})(k.q_{f})}\Big(\frac{1}{8}\tan
^{4}(\eta /2)+\frac{5}{8}+\frac{1}{4}\tan ^{2}(\eta /2)\Big)
\nonumber \\
&+&\frac{a^{2}w}{c^{2}}\Big(\frac{Q_{f}}{(k.q_{i})}+\frac{Q_{i}}{(k.q_{f})}-(\frac{
Q_{i}}{(k.q_{i})}+\frac{Q_{f}}{(k.q_{f})})\nonumber\\
&\times&(1+\tan^2(\eta /2))/2\Big), \label{14}
\end{eqnarray}

\begin{eqnarray}
B&=&\frac{w^{2}}{2c^{2}}\Big(\frac{(a_{1}.q_{i})(a_{1}.q_{f})}{(k.q_{i})(k.q_{f})}+\frac{(a_{2}.q_{i})(a_{2}.q_{f})}{(k.q_{i})(k.q_{f})}\tan
^{2}(\eta /2)\Big)  \nonumber \\
&-&\Big\{\frac{a^{2}}{2}+\frac{(a^{2})^{2}w^{2}}{2c^{4}(k.q_{i})(k.q_{f})}-\frac{%
a^{2}}{4}\Big(\frac{(k.q_{f})}{(k.q_{i})}+\frac{(k.q_{i})}{(k.q_{f})}\Big)
\nonumber \\
&+&\frac{a^{2}w^{2}}{2c^{2}(k.q_{i})(k.q_{f})}(q_{i}.q_{f}-c^{2})-\frac{
a^{2}w}{2c^{2}}(Q_{f}-Q_{i})\nonumber \\
&\times&(\frac{1}{(k.q_{f})}-\frac{1}{(k.q_{i})})\Big\}(1+\tan^2
(\eta /2))/2 , \label{15}
\end{eqnarray}

\begin{eqnarray}
C &=&\frac{w^{2}}{c^{2}(k.q_{i})(k.q_{f})}\Big(\cos (2\phi
_{0})\Big\{(a_{1}.q_{i})(a_{1}.q_{f})\nonumber\\
&-&(a_{2}.q_{i})(a_{2}.q_{f})\tan ^{2}(\eta /2)\Big\}+\sin (2\phi
_{0})\Big\{(a_{2}.q_{i})(a_{1}.q_{f})\nonumber\\
&+&(a_{1}.q_{i})(a_{2}.q_{f})\Big\}\tan (\eta /2)\Big)
-\Big\{\frac{a^{2}}{4}(\frac{(k.q_{f})}{(k.q_{i})}
+\frac{(k.q_{i})}{(k.q_{f})})\nonumber\\
&-&\frac{a^{2}w^{2}}{2c^{2}(k.q_{i})(k.q_{f})}((q_{i}.q_{f})-c^{2})-\frac{(a^{2})^{2}w^{2}}{2c^{4}(k.q_{i})(k.q_{f})}\nonumber\\
&+&\frac{a^{2}w}{2c^{2}}(Q_{i}-Q_{f})(\frac{1}{(k.q_{i})}-\frac{1}{(k.q_{f})})-%
\frac{a^{2}}{2} \Big\}\nonumber\\ &\times&\cos (2\phi _{0})(\tan
^{2}(\eta /2)-1) ,\label{16}
\end{eqnarray}

\begin{eqnarray}
D &=&-\frac{c}{2}\Big(\frac{(k.q_{f})}{(k.q_{i})}(\AA
.q_{i})+\frac{(k.q_{i})}{(k.q_{f})}(\AA.q_{f})\Big) \nonumber \\
&+&\frac{w}{c}\Big(\frac{Q_{i}(\AA.q_{f})}{(k.q_{f})}+%
\frac{Q_{f}(\AA.q_{i})}{(k.q_{i})}\Big)+\frac{c}{2}((\AA.q_{i})+(\AA.q_{f}))\nonumber
\\ &-&\frac{a^{2}w^{2}}{4c^{3}(k.q_{i})(k.q_{f})}(\tan ^{2}(\eta
/2)-1)\nonumber\\
&\times&((%
\AA.q_{i})+(\AA.q_{f})), \label{17}
\end{eqnarray}

\begin{eqnarray}
E &=&\cos (2\phi _{0})(\tan ^{2}(\eta /2)-1)a^{2}w  \nonumber \\
&\times&\Big\{-\frac{(q_{i}.q_{f})w}{4c^{2}(k.q_{i})(k.q_{f})}+
\frac{1}{4c^{2}}(\frac{Q_{i}}{(k.q_{i})}+\frac{Q_{f}}{(k.q_{f})})
\nonumber \\
&-&\frac{a^{2}w}{8c^{4}(k.q_{i})(k.q_{f})}+\frac{w}{4(k.q_{i})(k.q_{f})}\nonumber\\
&-&\frac{a^{2}w}{%
8c^{4}(k.q_{i})(k.q_{f})}\tan ^{2}(\eta /2)\Big\}, \label{18}
\end{eqnarray}

\begin{equation}
F=(\tan ^{2}(\eta /2)-1)^{2}\Big(\frac{(a^{2})^{2}w^{2}}{%
32c^{4}(k.q_{i})(k.q_{f})}\Big),  \label{19}
\end{equation}

\begin{eqnarray}
G &=&(\tan ^{2}(\eta
/2)-1)\frac{a^{2}w^{2}}{8c^{3}(k.q_{i})(k.q_{f})}\nonumber \\ &&
\Big\{\cos (3\phi _{0})((a_{1}.q_{i})+(a_{1}.q_{f}))\nonumber \\
&+&\sin (3\phi_{0})((a_{2}.q_{i})+(a_{2}.q_{f}))\tan (\eta
/2)\Big\}, \label{20}
\end{eqnarray}

\begin{eqnarray}
H &=&(\tan ^{2}(\eta
/2)-1)\frac{a^{2}w^{2}}{8c^{3}(k.q_{i})(k.q_{f})} \nonumber \\
&&\Big\{\cos (\phi _{0})(a_{1}.q_{i}+a_{1}.q_{f})\nonumber \\
&-&\sin (\phi _{0})(a_{2}.q_{i}+a_{2}.q_{f})\tan (\eta /2)\Big\},
\label{21}
\end{eqnarray}
where $\AA=a_1\cos(\phi_0)+a_2\sin(\phi_0)\tan(\eta/2)$. In the
absence of the laser field, all the contributions coming from the
sum over $s$ of the various ordinary Bessel functions vanish
except for $s=0$ where $J_{s}(0)=\delta _{s0}$ and we recover the
well known formula for Mott scattering in the absence of the laser
field \cite{5}

\begin{equation}
\frac{d\overline{\sigma }}{d\Omega _{f}}=\frac{1}{4}\frac{Z^{2}\alpha ^{2}}{%
\mid \mathbf{p}\mid ^{2}\beta ^{2}}\frac{(1-\beta ^{2}\sin ^{2}(\theta /2))}{%
\sin ^{4}(\theta /2)},  \label{22}
\end{equation}
where $\theta =(\widehat{\mathbf{p}_{i},\mathbf{p}_{f}})$. It can
easily be checked that for the case of a linear polarization of
the laser field ($\eta =0$), the phase $\phi _{0}=0$ and the
results of S.M. Li \textit{et al} \cite{1} are straightforward to
obtain whereas for the circular polarization ($\eta =\pi /2$), we
find the results previously found by Y.\ Attaourti \textit{et al}
\cite{2}. Eq. (\ref{11}) is the relativistic generalization of the
Bunkin and Fedorov \cite{7} treatment and it contains the degree
of ellipticity of the laser field as a new key parameter.

\section{Results and discussion.}

In this section, we discuss the numerical simulations for the
differential cross sections of the Mott scattering by an
elliptically polarized laser field. We assume without loss of
generality that the target is a proton having a charge $Z=1$. The
$z$ axis is set along the direction of the field wave vector
$\mathbf{k}$, $a_{1}^{\mu }=(0,\mathbf{a}_{1})$ and $a_{2}^{\mu
}=(0,\mathbf{a}_{2})$ with the vectors $\mathbf{a}_{1}$ and
$\mathbf{a}_{2}$
such that $\mathbf{a}_{1}=\mid \mathbf{a}\mid (1,0,0)$ and $\mathbf{a}%
_{2}=\mid \mathbf{a}\mid (0,1,0)$ from which one deduces that $\overline{%
A^{2}}=a^{2}(1+(\tan (\eta /2))^{2})/2=-\mid \mathbf{a}\mid
^{2}(1+(\tan (\eta /2))^{2})/2$. Thus, as found by S.M. Li
\cite{1} for the linear polarization, we have
$\overline{A^{2}}=-\mid \mathbf{a}\mid ^{2}/2$ and for
the case of circular polarization \cite{2}, $\overline{A^{2}}=-\mid \mathbf{a}%
\mid ^{2}$. The laser frequency used is $w=0.043$ $(a.u)$ which
corresponds to a photon energy $\hbar w=1.17$ $eV$. The incident
electron kinetic energy is $T_{i}=2.7$ $keV$ which corresponds to
a relativistic parameter $\gamma =(1-\beta
^{2})^{-\frac{1}{2}}=1.0053$ and the electric field strength has
the value $\mathcal{E}=0.05$ $(a.u)=2.57$ $10^{8}$ $V/cm$. The
various differential cross sections are plotted as functions of
the angle $\theta _{f}$. For small scattering angles, typically
($1^{{{}^{\circ }}}\leq \theta
_{i}\leq 15^{{{}^{\circ }}},1^{{{}^{\circ }}}\leq \phi _{i}\leq 15^{{%
{}^{\circ }}}$), ($-180^{{{}^{\circ }}}\leq \theta _{f}\leq 180^{{{}^{\circ }%
}},\phi _{f}=\phi _{i}+90^{{{}^{\circ }}}$), the summed
spin-unpolarized
differential cross sections are sharply peaked around $\theta _{f}=0^{{%
{}^{\circ }}}$ and are all close to the corresponding unpolarized
laser-free differential cross section given in Eq. (\ref{22}).
\begin{figure}[ht]
\centering
\includegraphics[angle=0,width=4 in,height=4 in]{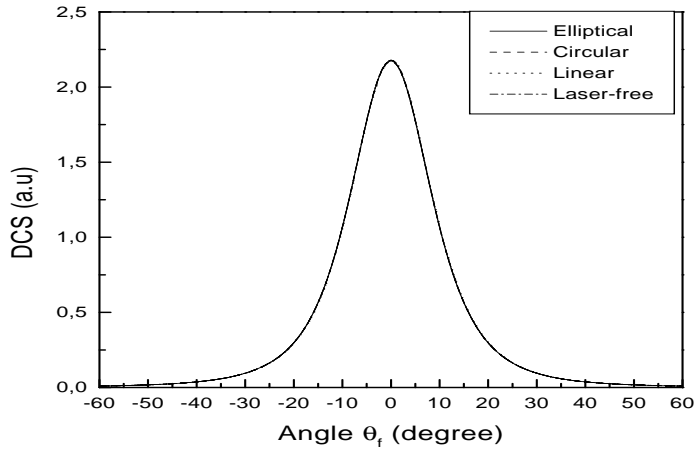}
\caption{The spin unpolarized summed cross sections for an
incident electron kinetic energy of $2.7$ $keV$ as a
function of the angle $\theta _{f}$ scaled in $10^{-2}$. The electric field strength is $%
\mathcal{E}=0.05$ $(a.u)$ and the laser frequency is $w=0.043$
$(a.u)$. The corresponding Mott-scattering geometry is given in
the text.}
\end{figure}
The three DCSs corresponding to $\eta =0$, $\eta =2\pi /3$ and
$\eta =\pi /2$
are given in Fig. 1 together with the laser-free DCS for the geometry ($%
\theta _{i}=15^{{{}^{\circ }}},\phi _{i}=15^{{{}^{\circ }}}$), ($-180^{{%
{}^{\circ }}}\leq \theta _{f}\leq 180^{{{}^{\circ }}},\phi
_{f}=\phi _{i}+90^{{{}^{\circ }}}$). We obtain four almost
indistinguishable curves. Thus, at small angles, the summed
differential cross sections are almost unmodified by the laser
field and its polarization does not play a key role. The physical
explanation of this observation is that classically, when the
particles are close to the small angle scattering region, this
corresponds to large impact parameters and the incident electron
does not deviate
notably from its trajectory. For other scattering angles, ($45^{{{}^{\circ }}%
}\leq \theta _{i}\leq 89^{{{}^{\circ }}},45^{{{}^{\circ }}}\leq
\phi _{i}\leq 89^{{{}^{\circ }}}$), ($-180^{{{}^{\circ }}}\leq
\theta _{f}\leq 180^{{{}^{\circ }}},\phi _{f}=\phi
_{i}+90^{{{}^{\circ }}}$), the situation changes drastically since
for medium and large scattering angles, the momentum transfer
during the Mott scattering is large and a significant number of
photons can be exchanged with the laser field.

In Fig. 2, we compare the three DCSs corresponding to $\eta=0$,
$\eta =2\pi /3$
and $\eta =\pi /2$ together with the laser-free DCS for the geometry ($%
\theta _{i}=60^{{{}^{\circ }}},\phi _{i}=0^{{{}^{\circ }}}$), ($-180^{{%
{}^{\circ }}}\leq \theta _{f}\leq 180^{{{}^{\circ }}},\phi _{f}=90^{{%
{}^{\circ }}}$) for an exchange of $\pm 150$ photons. In this
geometry, the effect of the laser field polarization is clearly
shown since the three DCSs are now well distinguishable. One has
to sum over a very large number of photons to recover the
laser-free DCS. Furthermore, all numerical simulations have shown
the following. The DCS for linear polarization is always higher
than the two others. The DCS for elliptical polarization is lower
or higher than the DCS for circular polarization depending on the
value of the degree of ellipticity $\eta $.
\begin{figure}[ht]
\centering
\includegraphics[angle=0,width=5.8 in,height=3.5 in]{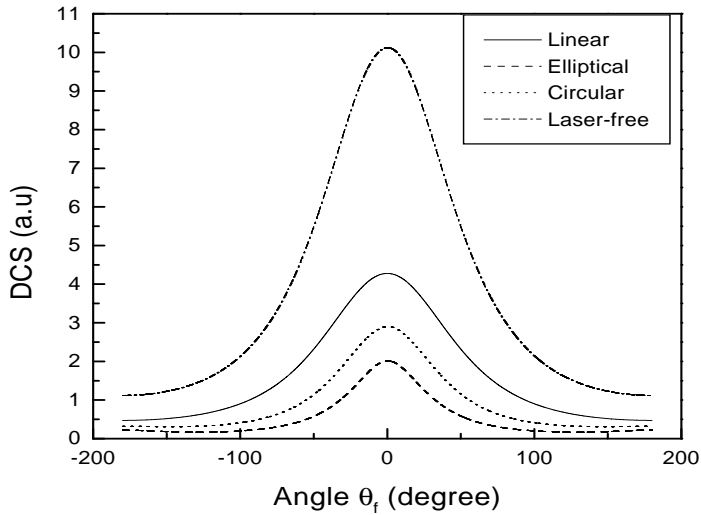}
\caption{The summed spin-unpolarized
cross sections for an exchange of $\pm 150$ photons scaled in $10^{-5}$. As in Fig. 1, $\mathcal{%
E}=0.05$ $(a.u)$ and $w=0.043$ $(a.u)$. The corresponding
Mott-scattering geometry is explained in the text.}
\end{figure}

To have an idea about the behaviour of the DCS as a function of
the degree of ellipticity $\eta $, we have obtained a three
dimensional curve corresponding to the same geometry as in Fig. 2
but for a degree of ellipticity $\eta $ varying from $0$ to $\pi
/2$. There are oscillations of the DCS for the elliptical
polarization and the DCS for linear polarization is always the
higher DCS. These oscillations decrease as the number of photons
exchanged is increased. The convergence towards the laser-free DCS
is faster for the linear polarization of the laser field. For the
circular polarization, this convergence is easily obtained when
one finds the value for the convergence corresponding to the
linear polarization whereas it is much more difficult to infer
from the previous results for which value of the number of photons
exchanged, the DCS for the elliptical polarization will converge
to the laser-free DCS. However, depending on the value of $\eta $
and in the non relativistic regime we have chosen, this number is
$\pm 1250$ photons. When the incident electron relativistic
parameter
 is increased from $\gamma =1.0053$ to $\gamma =(1-\beta ^{2})^{-\frac{1}{2}}=2$, the
previous mentioned results remain valid but the corresponding DCSs
are very small indicating a small probability that the very fast
projectile electron will exchange photons with the radiation
field. As for the behaviour of the DCSs with respect to the degree
of ellipticity, the ellipically and circularly polarized laser
modified cross sections become more sharply peaked around the
angle $\theta _{f}=0^{{{}^\circ}}$. A similar result has been
reported [3].

\section{Conclusions}

In this work, we have extended the study of the Mott scattering
process of an electron by a charged nucleus to the case of a
general polarization. We have shown that the Mott-scattering
geometry as well as the key parameters such as the electric field
strength and the incident electron kinetic energy
influence the behaviour of the DCSs. Moreover, the degree of ellipticity $%
\eta $ is also a key parameter for the description of the Mott
scattering process particularly in the region of large momentum
transfer and for a number of photons exchanged lower than that for
which the DCSs tend to the laser-free one.

\end{document}